\documentclass[aps,pra,twocolumn]{revtex4}

\usepackage{graphicx}

\begin{document}

\title{On the relation between transformation dynamics\\ and quantum statistics in weak measurements}

\author{Holger F. Hofmann}
\email{hofmann@hiroshima-u.ac.jp}
\affiliation{
Graduate School of Advanced Sciences of Matter, Hiroshima University,
Kagamiyama 1-3-1, Higashi Hiroshima 739-8530, Japan}
\affiliation{JST, CREST, Sanbancho 5, Chiyoda-ku, Tokyo 102-0075, Japan
}

\begin{abstract}
Experimentally, the imaginary parts of complex weak values are obtained from the response of the system to small unitary phase shifts generated by the target observable. The complex conditional probabilities obtained from weak measurements can therefore be explained in terms of transformation dynamics. Specifically, the complex phase of weak conditional probabilities provides a complete description of the transformation dynamics between the initial and the final state generated by the intermediate states. The result is a measure of quantum state overlap that relates quantum statistical properties directly to the dynamical action of unitary transformations.
\end{abstract}

\maketitle

\section{Introduction}

It is a well established fact that the statistical properties of quantum systems do not conform to classical expectations. Starting with the formulation of Bell's inequalities for the correlations between the properties of spatially separated systems, quantum paradoxes have been analyzed in terms of classical limits for the joint probabilities of measurements that cannot be performed jointly. In principle, quantum paradoxes can therefore be resolved by identifying the quantum mechanical equivalent of joint probabilities. Recently, such resolutions of quantum paradoxes have been achieved by using weak measurements \cite{Res04,Jor06,Wil08,Lun09,Yok09,Gog11}. By measuring the average of a weak interaction with the target observable, such measurements can avoid the back-action required by the uncertainty principle, allowing an experimental determination of correlations between the weak measurement observable and a final measurement. It is then possible to obtain the complete quantum statistics of a state conditioned by initial preparation and final measurement \cite{Hof10}. It is always possible to express these results in terms of a conditional probability distribution defined by a complete orthogonal basis ${\mid m \rangle}$ of Hilbert space. Weak measurements thus provide a natural definition of conditional and joint probabilities for measurements that cannot be performed jointly in quantum mechanics \cite{Hof10,Ste95,Hos10}.   

The possibility of negative weak conditional probabilities provides a convenient explanation for the statistical oddities of quantum paradoxes \cite{Res04,Jor06,Wil08,Lun09,Yok09,Gog11}. However, the sequence of preparation, weak measurement and post-selection also introduces a complex phase factor. To properly understand the physics of weak conditional probabilities, it is essential to explain the origin and the meaning of this complex phase. In this context, it is important to note that the imaginary part of a weak value is actually a response of the system to weak unitary transformations generated by the target observable \cite{Hof11a}.  It is then possible to understand the structure of weak conditional probabilities - and hence the fundamental structure of quantum statistics - in terms of the effects of unitary transformations on the transition between an initial and a final state \cite{Hof11b}. 

In the following, I will illustrate the relation between weak measurement statistics and dynamics using the example of a free particle given in \cite{Hof11b}. Relations with the classical action are considered and the implications for realistic and deterministic interpretations of physics are discussed. 

\section{Imaginary weak values}

If weak values are defined in terms of the average shift of a pointer observable, it is obvious that they should be given by real numbers. However, the mathematical form of the weak value of an observable $\hat{A}$ conditioned by an initial state $\mid i \rangle$ and a final state $\mid f \rangle$ is usually given by the complex form
\begin{equation}
\langle \hat{A} \rangle_{\mathrm{weak}} = \frac{\langle f \mid \hat{A} \mid i \rangle}{\langle f \mid i \rangle}.
\end{equation}
The pointer shift observed in a weak measurement is determined by the real part of the weak value. On the other hand, the imaginary part of the weak value is observed when $\hat{A}$ is the generator of a weak unitary transformation acting on the system \cite{Hof11a,Hof11b}. For $\hat{U}(\phi)=\exp(-i \phi \hat{A})$, the imaginary weak value is
\begin{equation}
\label{eq:logderiv}
\mathrm{Im}\left(
\langle \hat{A} \rangle_{\mathrm{weak}} 
\right)=\frac{1}{2}\frac{\partial}{\partial \phi} \ln \left(p(f|i)\right)\bigg|_{\phi=0}.
\end{equation}
Weak values thus combine the measurement statistics of an observable with the dynamic responses generated by the same observable. In particular, the interpretation of real weak values in terms of quantum statistics with negative probabilities can then be extended to include an interpretation of imaginary conditional probabilities as statistical weights of the dynamic response \cite{Hof11b},
\begin{equation}
\mathrm{Im}\left(\langle \hat{A} \rangle_{\mathrm{weak}} 
\right)=\sum_m A_m \; \mathrm{Im}\left(p(m|if)\right),
\end{equation}
where the weak conditional probability $p(m|if)$ of the eigenstate $\mid m \rangle$ is given by 
\begin{equation}
\label{eq:weakcond}
p(m|if)=\frac{\langle f \mid m \rangle \langle m \mid i \rangle}{\langle f \mid i \rangle}.
\end{equation}
Interestingly, the imaginary part of this conditional probability does not contribute at all to the correlations observed in quantum paradoxes, since these paradoxes do not depend on the sequence in which $m$ and $f$ are measured. It is therefore possible to omit the imaginary part in the definition of joint probabilities \cite{Hof10}. However, the opposite is not true - The response of the system to strong unitary transformations generated by $\hat{A}$ does depend on the real parts of the weak conditional probabilities of the eigenstates $\mid m \rangle$. 

\section{Strong unitary transformations}

The effect of the unitary operation $\hat{U}(\phi)$ can be represented by the phase shifts $\phi A_m$ applied to the phases of the eigenstate components $\mid m \rangle$. The effect of this transformation on the weak conditional probabilities of $m$ can be expressed in terms of this phase shift and a subsequent update of the normalization,
\begin{equation}
\label{eq:shift}
p(m|if;\phi) = \frac{\langle f \mid m \rangle\langle m \mid i \rangle}{\langle f \mid \hat{U}(\phi) \mid i \rangle} \exp\left(- i \phi A_m\right).
\end{equation}
Since the absolute square of the normalization $\langle f \mid \hat{U}(\phi) \mid i \rangle$ is the total probability $p(f;\phi)$ of finding the final result $f$ in the output, the dependence of this probability on the parameter $\phi$ can be derived by comparing the conditional probabilities at $\phi$ with the conditional probabilities at $0$. The result reads \cite{Hof11b}  
\begin{equation}
\label{eq:transcorr}
p(f;\phi) = \left|\sum_m \exp(-i \phi A_m)
p(m|if;0)
\right|^2 p(f;0).
\end{equation}
Unitary transformations with eigenstates $\mid m \rangle$ therefore modify the total output probability by either reducing or increasing the phase differences between the weak conditional probabilities. 

Although the relation between statistics and dynamics expressed by weak conditional probabilities has no classical analogy, it is possible to interpret the dynamics of the unitary in terms of canonical transformations generated by the action $\phi A_m$. Classically, the gradient of $\phi A_m$ in $m$ would determine the phase space distance corresponding to the shift of $\phi$ in the variable conjugate to $\hat{A}$. Therefore, the phase gradient of weak conditional probabilities corresponds to a phase space distance between the initial and the final state at a fixed value of $m$. Unitaries modify this distance by reducing or increasing this distance, pushing the state $\mid i \rangle$ towards or away from the final state $\mid f \rangle$. For a better visualization of these effects, it may be useful to consider the case of a particle moving in free space. 

\section{Trajectory of a kicked particle}

Let us consider the case of a particle moving from a position of $x(-\tau/2)=0$ to a position of $x(\tau/2)=0$. If we assume a constant velocity $v_y$ along the $y$-direction, this corresponds to the movement of a particle between two slits at a distance of $v_y \tau$ from each other. Classically, the particle must move in a straight line and is expected to be at $x(0)=0$ in the middle between the two slits. However, quantum mechanics defines the initial and the final states separately, as position eigenstates at $t_i=-\tau/2$ and at $t_f=\tau/2$, respectively. To reach a position $x$ at $t_m=0$ from its initial position, the particle should have a transverse momentum of $P_i=2m x/\tau$. Likewise, a particle from position $x$ at $t_m=0$ requires a momentum of $P_f=-2m x/\tau$ to reach the final position. Therefore, the initial and the final state can be represented by eigenstates of the corresponding linear combinations of position and momentum operators at $t_m=0$,
\begin{eqnarray}
\label{eq:eigen}
\left(\hat{x}-\frac{\tau}{2 m} \hat{P}\right) \mid i \rangle &=& 0 
\nonumber \\
\left(\hat{x}+\frac{\tau}{2 m} \hat{P}\right) \mid f \rangle &=& 0.
\end{eqnarray}
Note that quantum mechanics appears to confirm classical determinism at this point: For each individual state, the time evolution merely converts position information into a combination of position and momentum information according to the classical laws of motion. 

The quantum states that solve the eigenvalue equations (\ref{eq:eigen}) can be expressed in the position basis at $t_m=0$ as 
\begin{eqnarray}
\langle x \mid i \rangle &=& \frac{1}{\sqrt{L}} \exp\left(i \frac{m}{\hbar \tau} x^2 \right)
\nonumber \\
\langle x \mid f \rangle &=& \frac{1}{\sqrt{L}} \exp\left(- i \frac{m}{\hbar \tau} x^2\right),
\end{eqnarray}
where $L$ is a normalization length that should be sufficiently large to cover the relevant range of $x$-values. The weak conditional probability of $x$ obtained from these states is given by
\begin{equation}
\label{eq:xif}
p(x|if) = \sqrt{\frac{2 m}{\pi \hbar \tau}} \exp \left(i \frac{2 m}{\hbar \tau} x^2 - i \frac{\pi}{4}\right).
\end{equation} 
The real part of this probability is positive for $x^2 < 3 \pi \hbar \tau/(8 m)$. The normalization confirms that these positive contributions are sufficient to explain the total probability. The complex contributions from $x^2 > \hbar \tau/m$ average out to zero, leaving no net contributions from $x$-values far away from the classical result of $x=0$. 

Classically, a particle going from $x(t_i)=0$ to $x(t_f)=0$ through an intermediate position of $x$ at $t=0$ needs to be kicked at $t=0$ to compensate for the difference in momentum between the incoming state $i$ and the outgoing state $f$. The necessary change in momentum is $\Delta P = 4 mx/\tau$, corresponding to the phase gradient of the weak conditional probability,
\begin{equation}
\Delta P = \hbar \frac{\partial}{\partial x} \mathrm{Arg}\left(p(x|if)\right).
\end{equation}
Since position is the generator of momentum shifts, the canonical transformation can be expressed in terms of the position dependent action
\begin{equation}
S(x)=\hbar \mathrm{Arg}\left(p(x|if)\right).
\end{equation}
Therefore, the complex phase of the weak conditional probability corresponds to the action of a canonical transformation that corrects the momentum difference between the incoming particle and the outgoing trajectory to the target at $x(t_f)=0$. 

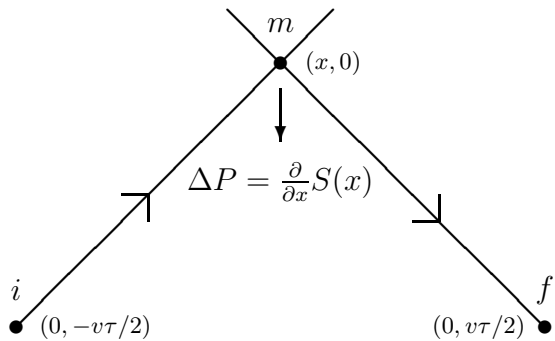
\begin{figure}[th]
\begin{picture}(240,150)
\thicklines
\put(20,20){\line(1,1){120}}
\put(220,20){\line(-1,1){120}}
\put(70,70){\line(-1,0){10}}
\put(70,70){\line(0,-1){10}}
\put(180,60){\line(-1,0){10}}
\put(180,60){\line(0,1){10}}

\put(25,10){\makebox(50,20){$(0,-v\tau/2)$}}
\put(120,110){\makebox(40,20){$(x,0)$}}
\put(170,10){\makebox(50,20){$(0,v \tau/2)$}}

\put(20,20){\circle*{5}}
\put(120,120){\circle*{5}}
\put(220,20){\circle*{5}}

\put(10,24){\makebox(20,20){\large $i$}}
\put(110,124){\makebox(20,20){\large $m$}}
\put(210,24){\makebox(20,20){\large $f$}}

\put(120,110){\vector(0,-1){20}}
\put(70,65){\makebox(100,20){\large $\Delta P = \frac{\partial}{\partial x} S(x)$}}

\end{picture}
\caption{\label{fig1} Schematic illustration of the change of momentum needed to kick the particle from $x(t_i)=0$ to $x(t_f)=0$. The action $S(x)$ is equal to $\hbar$ times the complex phase of the weak conditional probability $p(x|if)$.}

\end{figure}

Figure \ref{fig1} illustrates this identification of the complex phase of the weak conditional probability $p(x|if)$ with the action required to correct the trajectory from initial state $i$ to final state $f$ at an intermediate point $m$. By applying a unitary transformation $\hat{U}_{\mathrm{max}}=\exp(-i S(\hat{x}))$ at $t=0$, the transition probability from $i$ to $f$ is maximized to $|\langle f \mid \hat{U}_{\mathrm{max}} \mid i \rangle|^2=1$. The complex phase of the weak conditional probability thus defines the action of the unitary transformation that maximizes the transition probability from $i$ to $f$ \cite{Hof11b}. 

Oppositely, it may be interesting to consider the effect of a momentum kick of $\Delta P$ on the weak conditional probability. According to Eq. (\ref{eq:shift}), the corresponding unitary transform $\hat{U}(\Delta P)=\exp(-i \Delta P \;\hat{x}/\hbar)$ changes the weak conditional probability to
\begin{eqnarray}
\label{eq:xifmod}
\lefteqn{p(x|if;\Delta P) =}
\nonumber \\ && \sqrt{\frac{2 m}{\pi \hbar \tau}} \exp \left(i \frac{2 m}{\hbar \tau} (x-\frac{\Delta P\; \tau}{4 m})^2 - i \frac{\pi}{4}\right).
\end{eqnarray} 
The weak conditional probability is now centered around the classical result of $x=\Delta P\; \tau/(4 m)$ obtained for a kick of $\Delta P$ between $i$ and $f$. Thus the effect of a force on the trajectory is approximately reproduced in quantum mechanics. However, the precise deterministic relations are replaced by complex conditional probabilities, where the classical distance between the initial and the final state at the intermediate measurement result is represented by complex oscillations of the conditional probability. 

\section{Action as a measure of logical tension}

In the specific case of particle trajectories, the phase gradient of the weak conditional probability can be interpreted as a momentum difference between the initial and the final states. In general, it may often be difficult to evaluate the differences between the states in terms of their physical properties. Nevertheless, the phase of the weak conditional probabilities can serve as an indicator of agreement or disagreement between initial, intermediate and final states. In particular, quantum paradoxes originate from phases larger than $\pi/2$, which are only possible if the unitary operation associated with the weak conditional probabilities is rather strong. In this sense, the phase of the weak conditional probability can serve as a fundamental measure of the logical tension between a set of three states. In general, the logical tension of three states, $\mid i \rangle$, $\mid m \rangle$ and $\mid f \rangle$ is given by \cite{Hof11b}
\begin{equation}
S(i,m,f) = \mathrm{Arg}\left(\langle f \mid m \rangle \langle m \mid i \rangle \langle i \mid f \rangle \right).
\end{equation}
This value is unchanged under permutations of the states, so it applies equally to probabilities of $f$ conditioned by $m$ and $i$ or to probabilities of $i$ conditioned by $f$ and $m$. It is therefore possible to apply it as the action of unitaries maximizing the transition from $m$ to $i$ in $f$, from $f$ to $m$ in $i$, or from $i$ to $f$ in $m$. 

As the example from particle trajectories has shown, the physical properties associated with $i$, $m$ and $f$ usually contradict each other. Specifically, classical physics would define a functional relation $m(i,f)$, so that $i$ and $f$ would determine $m$. Likewise, $m$ and $i$ can determine $f(m,i)$ and $f$ and $m$ can determine $i(f,m)$. In quantum mechanics, complex conditional probabilities are obtained even though the functional relations between the properties are not fulfilled. Instead, the complex phase encodes the difference between the three pairs of conditions in terms of the action needed to move e.g. from $i$ to $f$ along a line of constant $m$. The classical relation $m(i,f)$ defines the value of $m$ that minimizes the action, so that results in the vicinity of $m(i,f)$ can contribute a slowly varying positive real part to the total probability. Far away from the classical solution, rapid oscillation of phase ensure that the contributions to the overall probability cancel. However, the precise statistical pattern now includes both positive and negative real parts, indicating a statistical relation between the observed outcome $f$ and outcomes observed after applying unitary transformations between $i$ and $f$. In this sense, weak conditional probabilities provide additional information about the logical relations between the three conditions, $i$, $m$ and $f$. By defining the phase of the weak conditional probabilities as the logical tension between the states, it is possible to express the non-classical nature of this logical relation between states that cannot be measured at the same time. 

\section{Quantum determinism}

Since phases ( or logical tensions) larger than $\pi/2$ can be used to explain quantum paradoxes, it is interesting to consider the implications of complex weak probabilities for the interpretation of quantum mechanics. At first sight, it may seem that weak probabilities are non-deterministic because they allow non-zero values of probability for intermediate results $m$ that do not fulfill the classical relation $m(i,f)$. However, the intermediate result can only be used in statistical predictions. Each individual system is fully determined by the initial and final conditions. It is therefore necessary to judge the information content of the non-classical relation between $i$, $f$ and $m$ in statistical terms. Here, it appears that the weak measurements realize a form of determinism by reducing the average uncertainties of all observables to zero \cite{Hos10,Hof11a}. In this sense, weak measurement statistics appear to replace classical determinism with a new kind of quantum determinism, where the ability to predict arbitrarily strong unitary transformations replaces the ability to attribute precise values to unobserved properties of the system. Weak measurement statistics thus suggest that realism can be abandoned without giving up determinism. Determinism plays an important role in the relations between cause and effect that are needed to analyze experiments, whereas the assumption of a non-empirical reality seems to have no such function. It is therefore possible to base a consistent interpretation of empirical reality on the assumption that the unobserved variables $m$ have no individual reality, but merely establish the statistical relations between different measurements that cannot be performed at the same time. Weak measurement statistics then ensure the consistency of these statistical relations with the fully deterministic causality relations of fundamental physics.  

\section{Conclusions}

Weak measurements suggest that quantum mechanics can be analyzed and understood in terms of complex conditional probabilities for the outcomes of alternative measurements. The role of the real parts of these conditional probabilities corresponds to the role of conditional probabilities in conventional Bayesian statistics, with the added possibility of negative values. However, the deeper reason for the complex values of probabilities is only revealed when transformations generated by the unobserved observables are considered. The imaginary part of conditional probabilities then describes the linear response to weak unitary transformations, while the complex phase carries the complete information on how the transition between initial and final state will respond to arbitrary transformations conserving the target observables of the weak measurement. 

In classical physics, a transformation changes the physical property of the system. The complex phases of weak conditional probabilities therefore indicate differences in the properties associated with the initial state $i$ and the final state $f$ at the intermediate state $m$. In terms of phase space contours, the crossing points of $i$ with $m$ and of $m$ with $f$ are separated by a distance equal to the phase gradient of the weak conditional probabilities. In classical physics, this would be a contradiction between the statements $i$, $m$ and $f$, and the conditional probability would be zero. Quantum mechanics allows a non-zero value of the probability by replacing the classical concept of agreement or contradiction with the complex phase of the conditional probability. The logical relation between three quantum states is therefore described by a combination of logical overlap given by the amplitude and logical tension given by the phase. Specifically, the logical tension established a non-classical correlation between the effects of unitary transformations and the statistical properties of quantum states that could explain quantum paradoxes as a necessary consequence of quantum determinism.

\end{document}